\documentclass[
reprint,
 amsmath,amssymb,
 aps,
]{revtex4-2}

\usepackage{graphicx}
\usepackage{dcolumn}
\usepackage{bm}

\begin{document}

\preprint{APS/123-QED}

\title{Eclipse avoidance in TianQin orbit selection}


\author{Bobing Ye}
\author{Xuefeng Zhang}
 \email{zhangxf38@sysu.edu.cn}
\author{Yanwei Ding}
\author{Yunhe Meng}
\affiliation{
 TianQin Research Center for Gravitational Physics and School of Physics and Astronomy, \\
 Sun Yat-sen University (Zhuhai Campus), Zhuhai 519082, P.R. China
}


\date{\today}

\begin{abstract}
In future geocentric space-based gravitational-wave observatory missions, eclipses due to passing through the Moon's and Earth's shadows can negatively impact the sciencecraft's thermal stability and steady power supply. The occurrence should be reduced as much as possible in orbit design. In regard to TianQin's circular high orbits, we tackle the combined challenges of avoiding eclipses and stabilizing the nearly equilateral-triangle constellation. Two strategies are proposed, including initial phase selection and orbit resizing to 1:8 synodic resonance with the Moon, where the latter involves slightly raising TianQin's preliminary orbital radius of $1\times 10^5$ km to $\sim 100900$ km. As the result, we have identified pure-gravity target orbits with a permitted initial phase range of $\sim 15^\circ$, which can maintain eclipse-free during the 3+3 month observation windows throughout a 5-year mission started in 2034, and meanwhile fulfil the constellation stability requirements. Thereby the eclipse issue for TianQin can be largely resolved. 
\end{abstract}

\maketitle



\section{\label{sec:level1} Introduction}

\begin{figure}[h]
\includegraphics[width=0.45\textwidth]{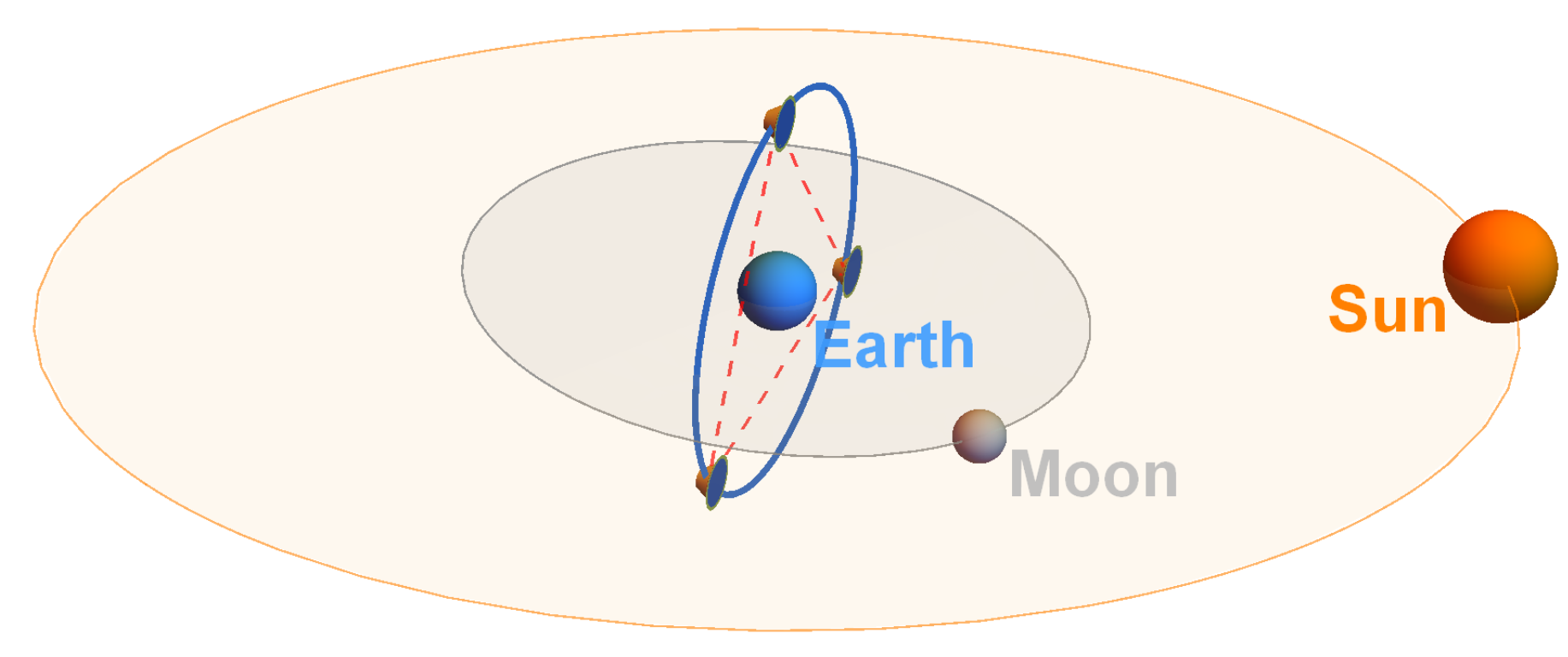}
\caption{\label{fig:EarthMoonSun} Illustration of the Sun, the Moon, and TianQin's orbits in an Earth-centered reference frame (not to scale). The lunar orbital plane is tilted from the ecliptic plane by $\sim 5^\circ$. }
\end{figure}

TianQin is a future space-based gravitational-wave (GW) observatory mission \cite{Luo2016} featuring circular high Earth orbits of a $10^5$ km radius, and a constellation plane nearly vertical to the ecliptic and facing the white-dwarf binary RX J0806.3$\texttt{+} $1527 (hereafter J0806) as a reference source. The mission engenders rich science prospects for GW physics and astronomy \cite{Hu2017,Huang2020,Liu2020,Wang2019,Feng2019,Fan2020,Shi2019,Bao2019}. The geocentric concept has benefits in launch cost, transfer duration, communication, telemetry, guidance and navigation, etc. One major challenge is the varying sunlight direction relative to the orbital plane (the beta angle), which affects the sciencecraft's thermal stability and the interferometric laser links (see Fig. \ref{fig:EarthMoonSun}). 
To deal with the issue, a preliminary 3+3 month operation scheme was suggested \cite{Luo2016}, in which one puts the GW observation on hold (for 3 months, to be optimized) when, twice a year, the Sun shines across the orbital plane from sideways. Moreover, a LISA-like passive thermal design concept adapted for the TianQin satellites was proposed \cite{Zhang2018,Zhang2019} as a potential candidate. Similar to other space observatories such as Gaia \cite{GAIA}, Wilkinson Microwave Anisotropy Probe (WMAP) \cite{WMAP}, and James Webb Space Telescope (JWST) \cite{JWST}, it adopts a flat-top sun-shield to keep critical devices protected in the shade (see Fig. \ref{fig:EarthMoonSun}), and shifts the thermal flux variation to a half-year period. The design implications are currently being evaluated \cite{Zheng2020,Chen2020}. Other recent progresses of TianQin can be seen in \cite{Mei2020}.

Related to the thermal issue, eclipses due to the Moon and the Earth's temporary blocking of sunlight has raised concerns for TianQin as well as for other geocentric concepts \cite{NASA2012}. First, passing through the shadows causes thermal disturbance to the sciencecraft, and in turn, to the sensitive science payloads mounted inside, where a core temperature stability of $\sim 10^{-5}$ K/Hz$^{1/2}$ in the mHz frequency band is typically needed \cite{Luo2016,LISA2017}. Second, eclipses interrupt the steady power output from the solar panels, which may also degrade the science payload performance. Concerning the hardware, it may be difficult and costly to design instruments to directly cope with the changing thermal environment during eclipses. Therefore, to ensure uninterrupted observations, it would be desirable that the satellites simply undergo \emph{no eclipses} at least during the 3+3 month observation windows within the 5-year mission lifetime. Here the 3+3 month observation windows correspond to when the sunlight direction makes an angle of $<45^\circ$ with the normal of the constellation plane (08 Jun to 06 Sep, 07 Dec to 07 Mar).

The problem of eclipse avoidance should be handled together with the constellation stability requirements in orbit design. Due to lunisolar gravitational perturbations and initial orbit errors, the constellation will drift away from the nominal equilateral triangle. The preliminary TianQin requirements on the constellation stability include the arm-length variations within $\pm 0.5\%$, the relative line-of-sight velocities between satellites within $\pm 10$ m/s, and the breathing angle variations within $\pm 0.2^\circ$ for 5 years \cite{Ye2019}. Therefore the challenges are two-fold: to minimize shadow events for all three satellites, and at the same time, to maintain the required level of the constellation stability.

The eclipse issue has been briefly discussed in other geocentric mission concepts \cite{NASA2012}. For instance, gLISA in geostationary orbits schedules switch-offs for 45 days during each Earth eclipse season in spring and fall \cite{Tinto2015}. In one-year period, Earth eclipses for LAGRANGE can last maximally 4 hours, and Moon eclipses 3 hours \cite{Buchman2012}. As for OMEGA, there can be long ($\sim 4$ hours), mostly penumbral, eclipses \cite{Hellings2012}. TianQin differs from these missions in both the orbital radius and orientation, and hence the eclipsing properties differ as well. Despite the issue having been pointed out quite early on, a remedy through orbit design appears to have not been well studied so far for geocentric concepts. Note that the heliocentric orbits of LISA \cite{LISA2017} are not subject to Earth and Moon eclipses since they are positioned quite far away ($\sim 5\times 10^7$ km, $\sim 20^\circ$ trailing angle from the Earth).

The practice of eclipse evasion is often conducted in space observatories, particularly those orbiting around the Sun-Earth Lagrange point L2, in order to protect sensitive instruments on-board. For examples, Herschel Space Observatory, Planck Space Observatory, Gaia Space Observatory, Wilkinson Microwave Anisotropy Probe (WMAP), James Webb Space Telescope (JWST) have all taken eclipses into account in their orbit design \cite{Hechler2003,Bauske2009,Cavaluzzi2008,Yu2019}. Specifically for Gaia, it has performed one eclipse avoidance manoeuvre through the mission control 6 years after the launch \cite{Gaia2019}. Another related example can be seen in Chang'e 4's relay satellite Queqiao near the Earth-Moon L2 point, which has adopted a halo orbit with carefully chosen amplitudes and phases in order to avoid eclipses and lunar occultation of the Earth \cite{Tang2017,Gao2017,Gao2019,Liang2015,Liang2019}. Useful eclipse avoidance strategies can also be found in the design of near rectilinear halo orbits \cite{Williams2017,Zimovan2017,Davis2020,McCarthy2019,Davis2017} and geostationary orbits \cite{Lundgren1970,Hart1992}.

This is the fourth paper of our concept study series on TianQin's orbit and constellation \cite{Ye2019,Tan2020,Zhang2020}. The first paper \cite{Ye2019} demonstrates the optimization of TianQin’s orbits and shows that the 5-year constellation stability requirements can be met. The second \cite{Tan2020} investigates the impact of orbital radius and orientation selections on the constellation stability, and the findings provide support to TianQin’s orbit design. The third \cite{Zhang2020} examines the disturbance from the Earth-Moon’s gravity to the intersatellite ranging measurements and points out that the effect present no showstopper to the mission. The current paper is structured as follows. In Section \ref{sec:eclipse}, eclipsing properties and statistics are discussed for TianQin. In Section \ref{sec:strategies}, three avoidance strategies are described, including avoidance maneuvers, initial phase selection, and orbit resizing. Sections \ref{sec:method} and \ref{sec:results} present the three-step optimization method and eclipse-free orbits found for the results. The concluding remarks are made in Section \ref{sec:conclusion}.


\section{\label{sec:eclipse} Eclipse Occurrence}

\begin{figure}[ht]
\includegraphics[width=0.45\textwidth]{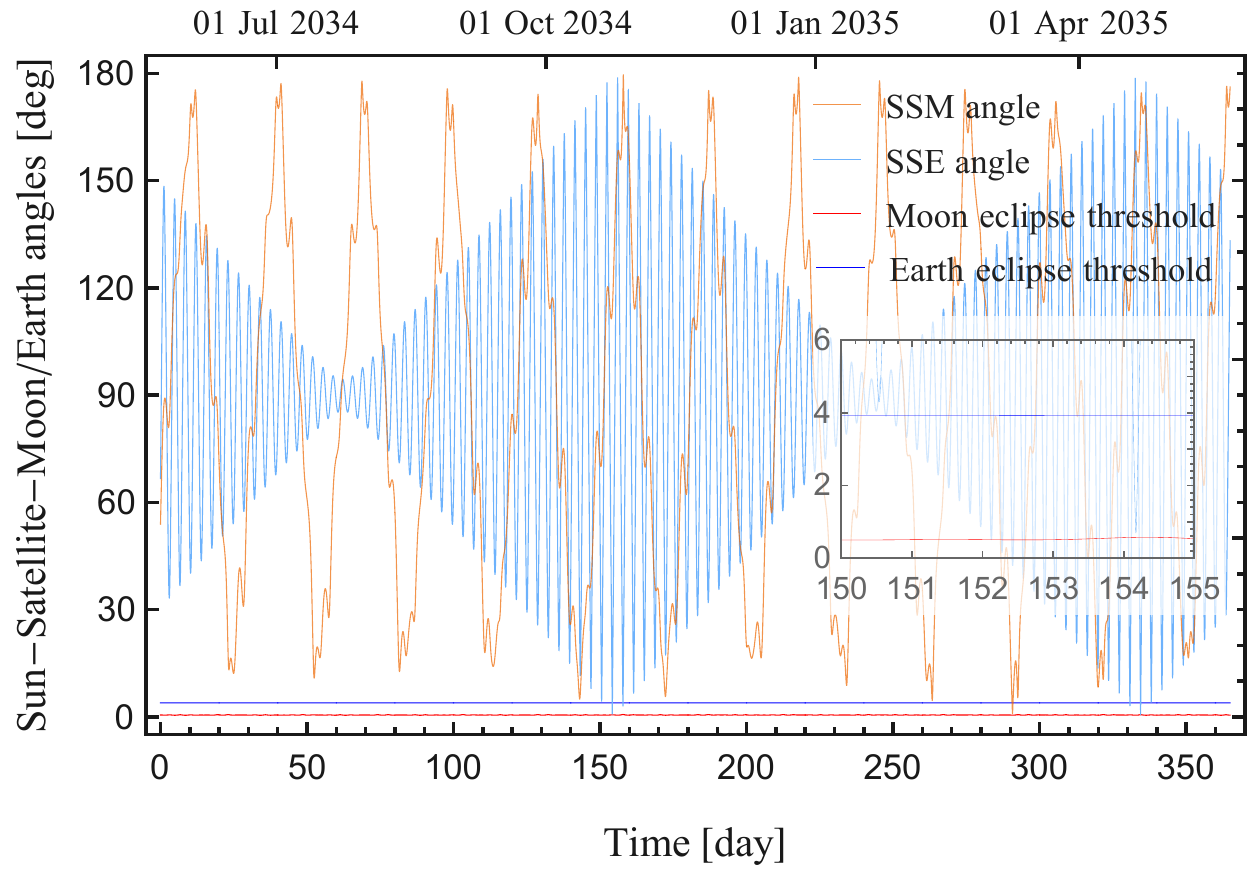}
\caption{\label{fig:SSM_SSE} An example of typical one-year evolution of the Sun-Satellite-Moon and Sun-Satellite-Earth angles for TianQin satellites. The thresholds for the Moon and Earth eclipses are marked, respectively, by red and blue lines, below which an eclipse takes place (e.g., day 154). }
\end{figure}

At the orbital radius of $10^5$ km, the TianQin satellites may experience eclipses due to the Moon and the Earth (see Fig. \ref{fig:EarthMoonSun}). The occurrences depend on their positions relative to the Sun. For TianQin's orbits, Fig. \ref{fig:SSM_SSE} shows an example of typical one-year evolution of the Sun-Satellite-Moon (SSM) and Sun-Satellite-Earth (SSE) angles. The initial orbital parameters are taken from Table 3 of \cite{Ye2019}, which were optimized without considering eclipse reduction. As the figure indicates, an eclipse takes places when the SSM and SSE angles are below the thresholds determined by the apparent sizes of the Sun ($\sim 0.53^\circ$ at 1 AU) and the occulting bodies \cite{Montenbruck2001}. In Table \ref{tab:eclipse_stat}, the statistics of predicted events are presented, and the data were obtained from \texttt{GMAT} \cite{GMAT}.

\begin{table}[ht]
\caption{\label{tab:eclipse_stat} The 5-year statistics of eclipse events for the three TianQin satellites before taking eclipse avoidance measures. The duration is in minutes. }
\begin{ruledtabular}
\begin{tabular}{ccccc}
Object & Time & Type & Number & Duration (mean) \\ 
\hline
Moon & $\pm 1.4$ days of & Partial\footnote{Penumbra only} & 17 & 24-70 (47) \\
     & new moons & Annular\footnote{Penumbra+antumbra+penumbra} & 1 & 57 \\
Earth & $\pm 5$ days of & Total\footnote{Penumbra+umbra+penumbra} & 57 & 60-114 (98) \\
      & 04/22 \& 10/24 & Partial & 8 & 16-53 (40) \\
\end{tabular}
\end{ruledtabular}
\end{table}

\begin{figure}[ht]
\includegraphics[width=0.45\textwidth]{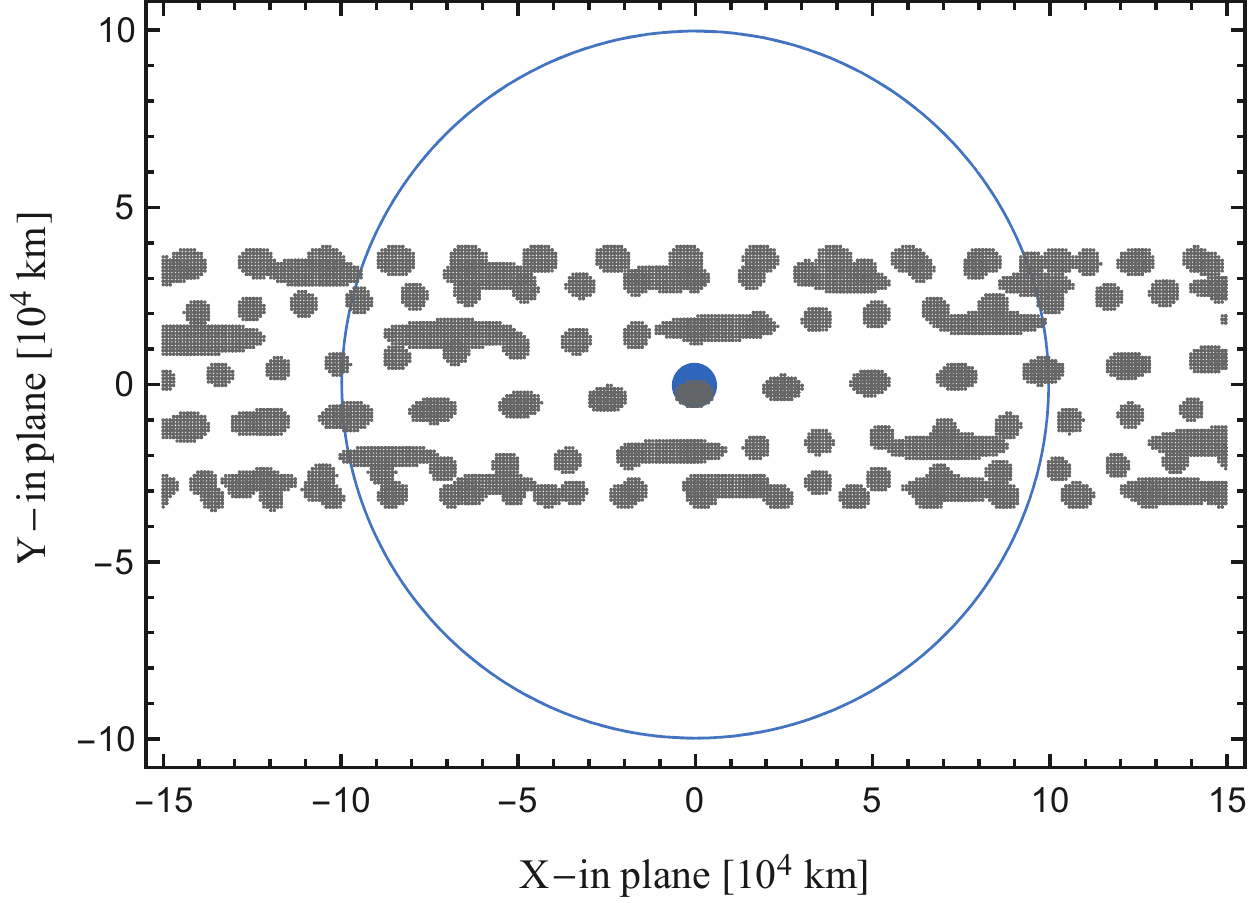}
\caption{\label{fig:proj} Intersection areas of the Moon's penumbra (shadow cone) and TianQin's orbital plane (X-Y plane), recorded every 0.2 day for one-year evolution. The X-axis is parallel to the ecliptic and the blue circle marks TianQin's orbit. }
\end{figure}

Moon eclipses may occur within 1.4 days before and after the new moon of every synodic month. The partial eclipses dominates in occurrences, and roughly half of them happen in the 3+3 month observation windows. The duration averages about 47 minutes. Furthermore, Fig. \ref{fig:proj} marks the intersection areas of the Moon's penumbra (shadow cone) with the TianQin's orbital plane every 0.2 day over the course of one year. At new moons, the shade sweeps horizontally through the orbital plane and leaves a trail of elliptical speckles over each passing. The vertical spreading of the trails is caused by the $\sim 5^\circ$ tilt of the Moon's orbital plane with respect to the ecliptic.

The situation of Earth eclipses is similar to that of geostationary orbits which has two eclipse seasons around the spring and autumnal equinoxes. For TianQin, the Earth eclipse seasons span $\pm 5$ days about 22 Apr. and 24 Oct., when sunlight aligns with the constellation plane. The eclipses last roughly 1.5 hours on average. As Fig. \ref{fig:SSM_SSE} indicates, falling in the Earth's shadow is unavoidable for the TianQin satellites (see also \cite{Chao1980}). But it only arises well outside the 3+3 month observation windows, and hence the related issues are less disconcerting. In this study, we will focus on reducing the Moon eclipses during 3+3 month observation windows within the 5-year mission lifetime.


\section{\label{sec:strategies} Eclipse Avoidance Strategies}

As mentioned earlier, the thermal stability requirement calls for the TianQin constellation not experiencing Moon eclipses during the 3+3 month observation windows (08 Jun to 06 Sep, 07 Dec to 07 Mar) throughout the nominal mission lifetime, and if possible, with extension to 4+4 months or longer. With TianQin's orbital plane facing J0806, there can be at least three strategies to achieve the goal.

1. \emph{Avoidance maneuvers}. This is to perform a planned maneuver prior to a predicted eclipse inside an observation window by firing on-board low-level thrusters during non-GW-observing periods. Propellant consumption is a major concern as it shortens the mission lifetime. Besides, maneuvering individual satellites must not undermine the configuration stability of the constellation. Hence, this is considered as a less favorable resort that may not be used frequently.

2. \emph{Initial phase selection}. It takes $\sim 1.8$ hours for the Moon's shadow to sweep across the orbital trajectory (see Fig. \ref{fig:proj}, the blue curve), which is much shorter than the TianQin's orbital period ($\sim 3.6$ days). Therefore it is possible to tune the initial phase angles of the satellites so as to steer clear of the Moon's shadow along pure-gravity orbits for an extended period of time. This also helps to reduce the number of ensuing avoidance maneuvers needed.

\begin{figure}[ht]
\includegraphics[width=0.45\textwidth]{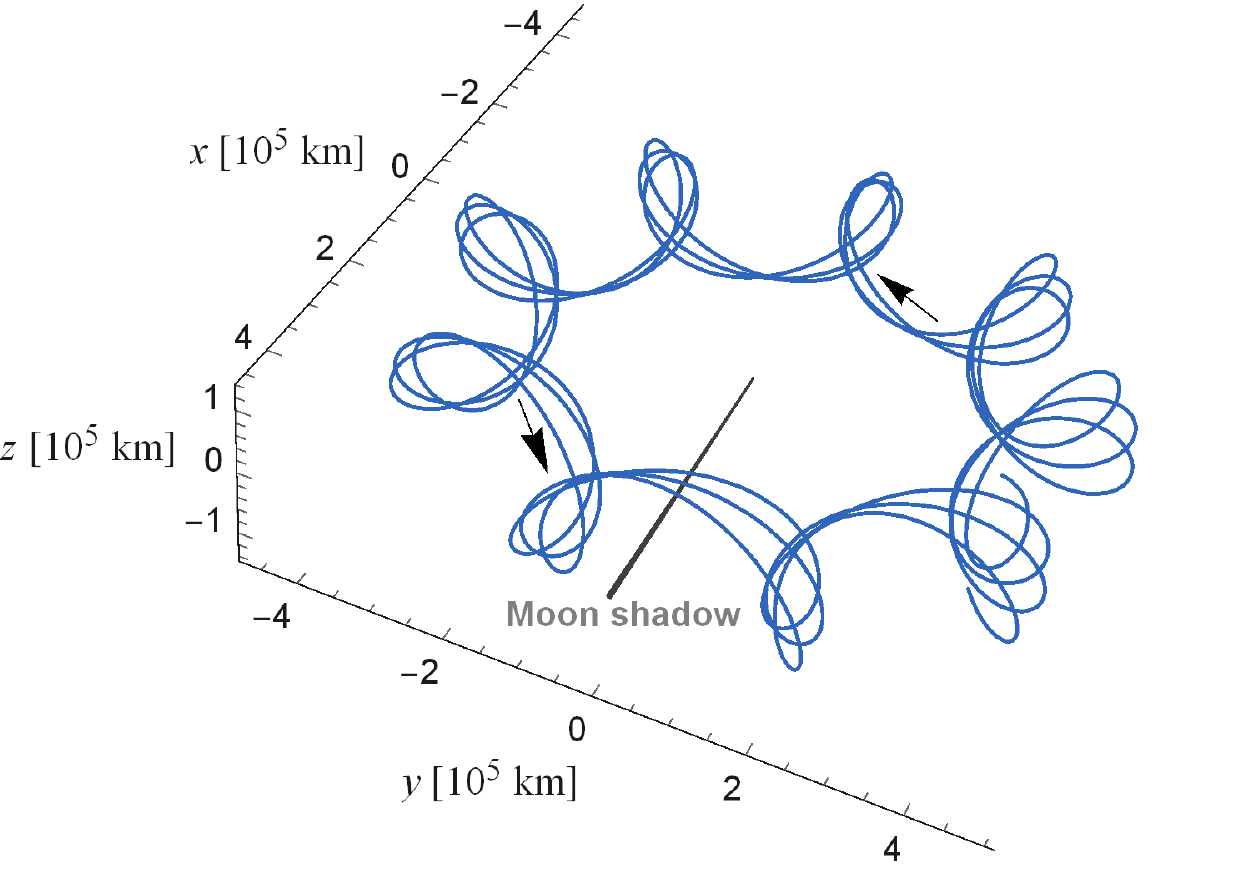}
\caption{\label{fig:TQ_traj} An example of 3-month trajectories of one TianQin satellite in 1:8 synodic resonance with the Moon, visualized in the Moon-centered coordinate system co-rotating with the Sun-Moon vector. The Moon's shadow cone, pointing along the positive $x$-axis, is aligned with a gap formed by the trajectory to avoid eclipses. For non-resonant orbits such as the $1\times 10^5$ km radius, the trajectory would appear more spread out with narrower gaps. }
\end{figure}

3. \emph{Orbit resizing}. This can be applied jointly with the second strategy to achieve an optimal performance. Particularly, by altering the orbit period (phase rate), one can set the motion of the satellites in a repeated phase relation with the motion of the Moon's shadow. Matching with the lunar phases, i.e., resonance in terms of synodic periods, can lower the possibility of crossing the Moon's shadow. Such a resonance can be visualized in the Moon-centered coordinate system co-rotating with the Sun-Moon vector (see Fig. \ref{fig:TQ_traj}). To avoid eclipses, the initial phase can be adjusted so that the Moon's shadow cone, projecting in the positive $x$-direction, passes through the large gaps formed by the repeated pattern of the trajectory.

\begin{figure}[ht]
\includegraphics[width=0.45\textwidth]{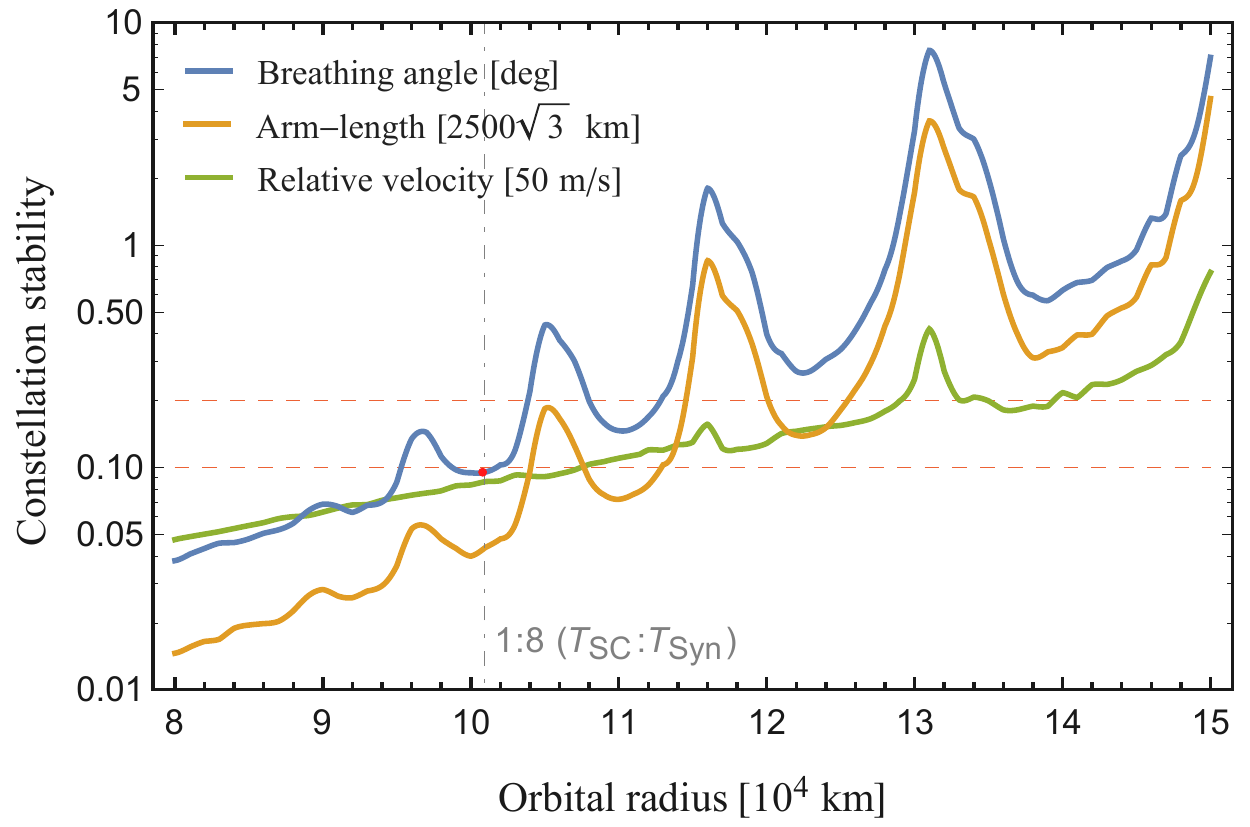}
\caption{\label{fig:syn-resonant} The dependence of TianQin's constellation stability on the orbital radius (mean semi-major axis) \cite{Tan2020}. The blue curve refers to the breathing angle variations, the yellow curve the arm-length variations, and the green curve the relative velocities. The upper red dashed line stands for the 5-year constellation stability requirements, and the lower red dashed line the requirements for the first 2 years \cite{Ye2019}. The 1:8 synodic resonance is marked by vertical dash-dotted line. }
\end{figure}

The preliminary TianQin design has picked an orbital radius of $1\times 10^5$ km \cite{Luo2016}. But the selection did not take into account evading Moon eclipses. Based on the strategies 2 and 3 above, we propose 1:8 synodic resonant orbits with regard to the Moon. This means to have the ratio of TianQin's orbital period and the synodic month (average 29.53 days) approximately 1:8. The choice elevates the orbital radius to about $1.009\times 10^5$ km which does not exceed the $1.13\times 10^5$ km limit set by the constellation stability requirements \cite{Tan2020}. Conveniently, the small increase in arm-length by $<1\%$ gives rise to negligible degradation in the constellation stability performance (deviations from the nominal equilateral triangle in terms of breathing angle and arm-length variations, and relative line-of-sight velocities between satellites, see Fig. \ref{fig:syn-resonant}), and bears no impact on expected science output and instrumentation. Though other options to mitigate eclipsing may also be considered, here we will attend to executing the second and third strategies for 1:8 synodic resonant orbits in the following sections.

\section{\label{sec:method} Optimization Method}

The orbit simulation setup follows our previous work \cite{Ye2019,Tan2020}. The force model includes the main solar system bodies as point masses as well as the Earth's non-spherical gravity field, and the propagation assumes pure-gravity orbits for the drag-free controlled satellites. 

The goal of optimizing TianQin's target orbit selection is to meet the constellation stability requirements and to be eclipse-free during each 3-month observation window (with possible extensions). The latter is mainly achieved by adjusting the orbital radii and initial phases within the stability constraint. The intensive search consists of 3 steps.

\emph{Step 1}. The initial epoch is set on 22 May, 2034 12:00:00 UTC, when the Earth-Moon vector is nearly perpendicular to the orbital plane. This helps to lower the eccentricity growth \cite{Tan2020} and the impact of initial phase adjustment on the constellation stability. Other epochs with such a property can also be used. 

\emph{Step 2}. We assume the parameter space of Table \ref{tab:par_space} for the initial orbital elements of the three satellites. The search region of the semi-major axis $a$ is extended to 100800-101000 km with $\Delta a = 5$ km sampling intervals to account for the length variation of synodic cycles (about 29.18 to 29.93 days). Since the three satellites are identical and forming an equilateral triangle, the initial phase of SC1 can be limited to 0-120$^\circ$ with a step size $\Delta\nu_{1} = 0.5^\circ$. 

\begin{table}[ht]
\caption{\label{tab:par_space} The search space for the initial orbital elements of the three TianQin satellites (SC1, 2, 3) in the J2000-based Earth-centered ecliptic coordinate system. $\nu_{1}$ denotes the true anomaly of SC1. }
\begin{ruledtabular}
\begin{tabular}{cccccccc}
$a$ & $e$ & $i$ & $\Omega$ &$\omega$ & $\nu_{1}$ & $\Delta a$ & $\Delta\nu_{1}$ \\
\hline
100\,800-101\,000 km & 0 & $94.7^\circ$ & $210.4^\circ$ &
$0^\circ$ & 0-120$^{\circ}$ & 5 km & $0.5^\circ$ \\
\end{tabular}
\end{ruledtabular}
\end{table}

\emph{Step 3}. Optimize each set of initial elements to stabilize the constellation using the efficient method of \cite{Tan2020} (see Sec. 3), and calculate resulting eclipse events. Then identify the range of orbital parameters that meets the eclipse-free condition.


\section{\label{sec:results} Results}

\begin{figure}[ht]
\includegraphics[width=0.45\textwidth]{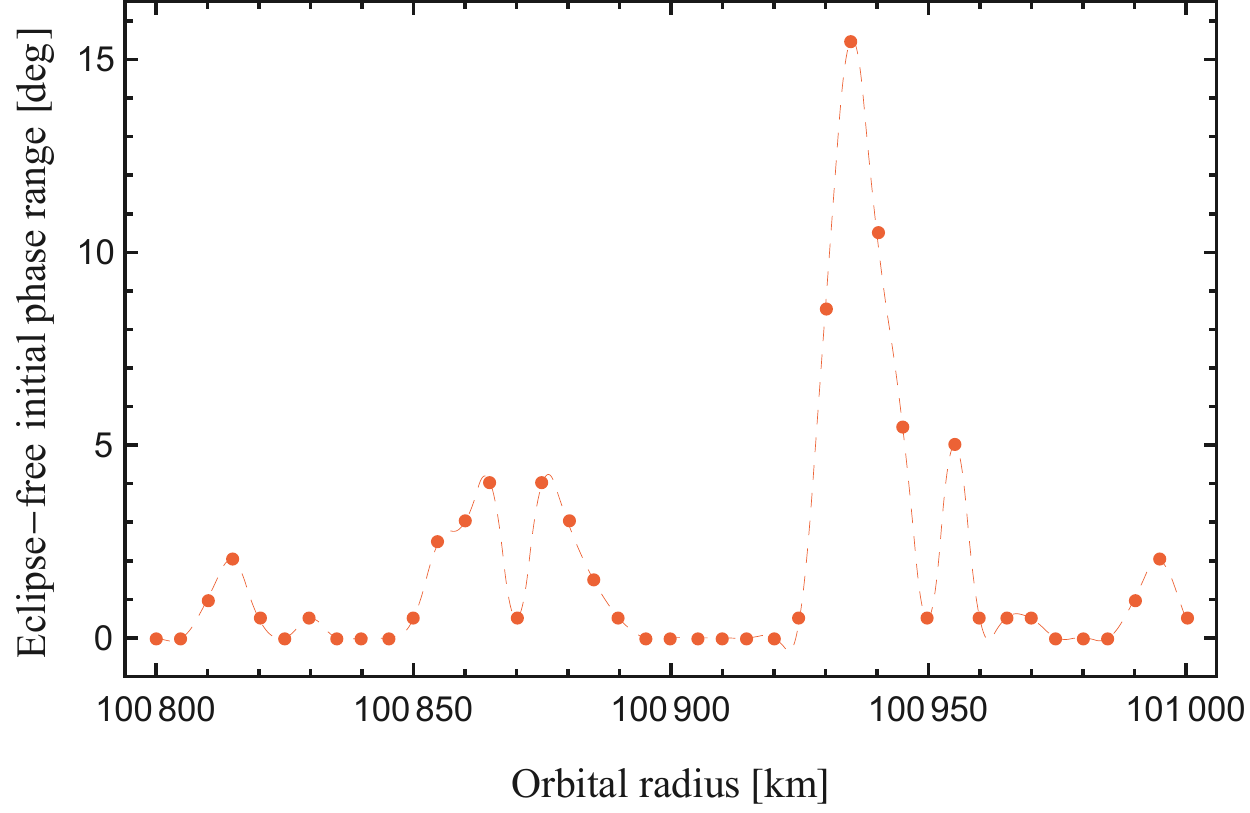}
\caption{\label{fig:phase_range} Eclipse-free initial phase ranges ($\nu_{\text{EF}}$) vs. orbital radii (mean semi-major axes). }
\end{figure}

Following the method above, Fig. \ref{fig:phase_range} shows the widths of the permitted eclipse-free initial phase ranges ($\nu_{\text{EF}}$, for all three satellites) at various orbital radii. The plot indicates that $a=100935$ km offers the widest range of initial phases for one to choose from. Note that a larger phase range would impose less requirements on initial orbit errors, hence easier to be realized. In contrast, for $a=10^{5}$ km, the method finds \emph{no orbits} that are eclipse-free during the 3+3 month observation windows for 5 years, starting from May 2034. Thus the 1:8 synodic resonant orbits are shown to perform much better than the non-resonant $a=10^{5}$ km orbits in avoiding eclipses. 

To examine more closely, a finer search with $\Delta a = 1$ km and $\Delta\nu_{1} = 0.1^\circ$ was carried out around $a=100935$ km, and the permitted initial phase ranges are summarized in Table \ref{tab:phase_range}. It can be seen that $a=100935$ km indeed corresponds to the largest $\nu_{\text{EF}}$, reaching $15.7^\circ$. Moreover, the eclipse-free periods can be extended at the cost of a reduced $\nu_{\text{EF}}$, for instance, to 6 years (May 2034 to May 2040) with $\nu_{\text{EF}}=15.6^\circ$, or, to 8 years (May 2034 to May 2042) and 4+4 month observation windows with $\nu_{\text{EF}} = 1.7^\circ$. 

\begin{table}[!ht]
\caption{\label{tab:phase_range} Eclipse avoidance results near $a=100935$ km with sampling intervals $\Delta a = 1$ km and $\Delta\nu_{1} = 0.1^\circ$, where $\nu_{1}$ denotes the permitted initial phases, and $\nu_{\text{EF}}$ the width of $\nu_{1}$. }
\begin{ruledtabular}
\begin{tabular}{ccc}
  $a$ &  $\nu_{1}$  & $\nu_{\text{EF}}$  \\ 
\hline
 100\,933 km & 73.6$ - $87.8$ ^\circ $& 14.2$ ^\circ $  \\ 
 100\,934 km & 74.0$ - $88.9$ ^\circ $& 14.9$ ^\circ $  \\
 100\,935 km & 74.3$ - $90.0$ ^\circ $& 15.7$ ^\circ $  \\
 100\,936 km & 75.8$ - $91.1$ ^\circ $& 15.3$ ^\circ $  \\
 100\,937 km & 77.2$ - $91.8$ ^\circ $& 14.6$ ^\circ $  \\
\end{tabular}
\end{ruledtabular}
\end{table}

\begin{figure}[ht]
\centering 
\begin{minipage}{0.36\textwidth}
\includegraphics[width=\textwidth]{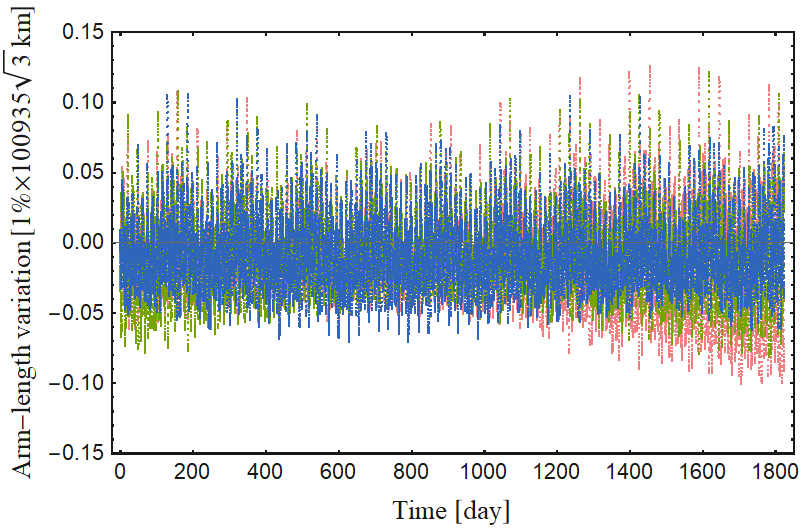}
\end{minipage}
\begin{minipage}{0.36\textwidth}
\includegraphics[width=\textwidth]{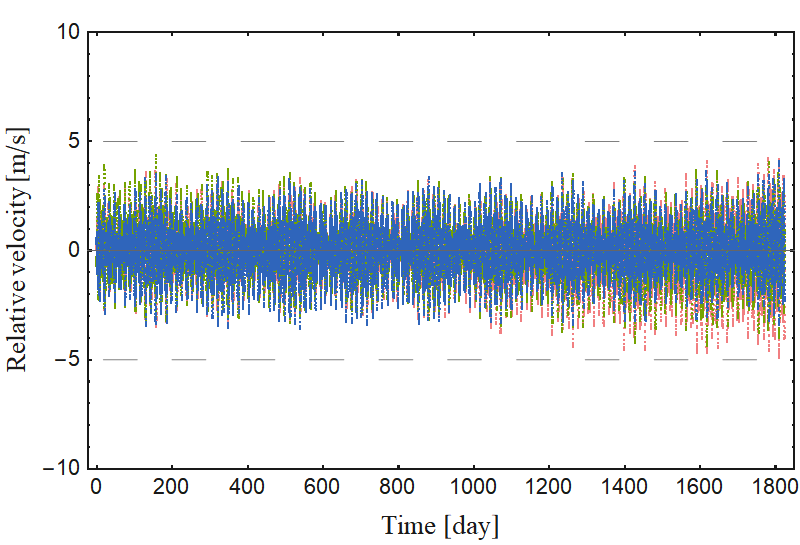}
\end{minipage}
\begin{minipage}{0.36\textwidth}
\includegraphics[width=\textwidth]{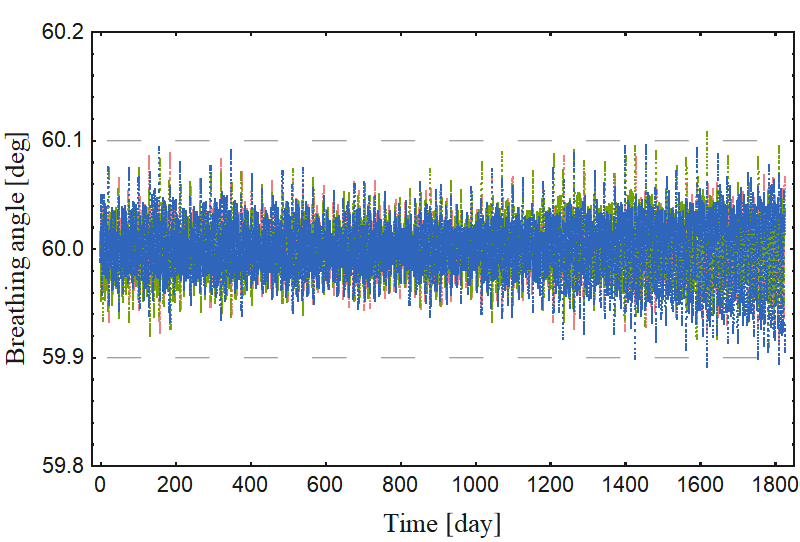}
\end{minipage}
\begin{minipage}{0.36\textwidth}
\includegraphics[width=\textwidth]{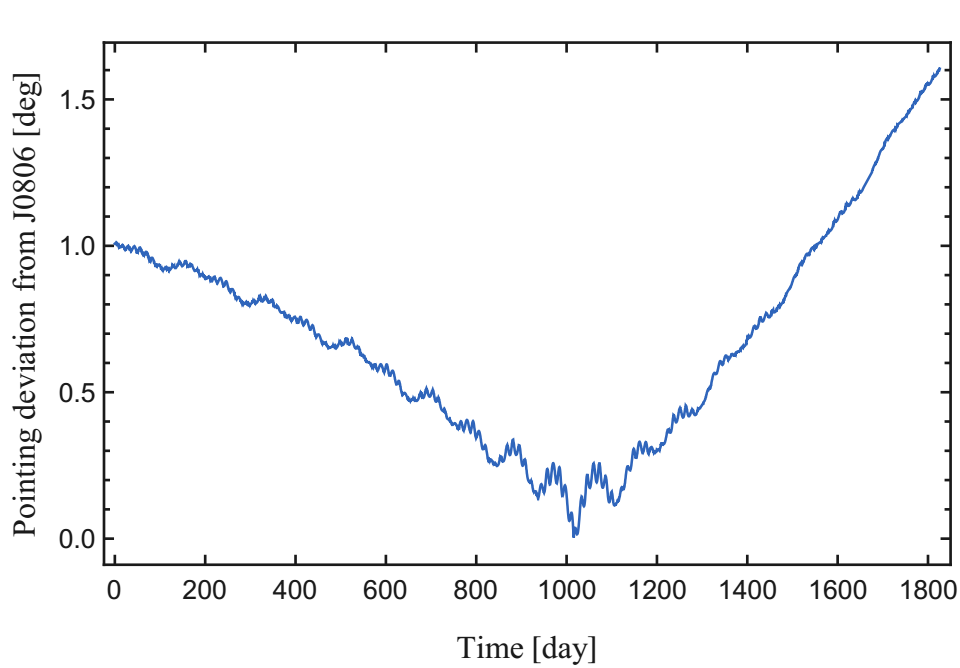}
\end{minipage}
\caption{\label{fig:stability} Evolution of the optimized TianQin orbits generated from the initial elements of Table \ref{tab:initial}. The plots show the 5-year variations of the arm-lengths, the relative velocities, the breathing angles, and the pointing deviation from J0806, respectively. The dashed lines correspond to the constellation stability requirements for the first 2 years \cite{Ye2019}. } 
\end{figure}

To demonstrate constellation stability, we take a set of eclipse-free optimized orbits with $\bar{a}=100935$ km as an example (see Table \ref{tab:initial} and Table \ref{tab:eclipse_lun}). Fig. \ref{fig:stability} shows that the 5-year evolutions of the arm-lengths, the relative velocities, and the breathing angles, indeed meet the constellation stability requirements. Likewise, the orbits with other permitted initial phases in Table \ref{tab:phase_range} can be optimized to the same level.

\begin{table}[ht]
\caption{\label{tab:initial} The initial elements of a set of optimized TianQin orbits in the J2000-based Earth-centered ecliptic coordinates at the epoch 22 May, 2034 12:00:00 UTC. The subsequent orbital evolution meets the 3+3 month eclipse-free and constellation stability requirements (see Table \ref{tab:eclipse_lun} and Fig. \ref{fig:stability}). }
\begin{ruledtabular}
\begin{tabular}{ccccccc}
 & $ a $\,(km) &  $ e $ & $ i $\,($ ^{\circ} $) \\ 
\hline
SC1 & 100\,926.158\,459 & 0.000\,300 & 94.774\,822  \\
SC2 & 100\,940.789\,023 & 0.000\,019 & 94.782\,183  \\
SC3 & 100\,938.056\,412 & 0.000\,411 & 94.785\,623  \\
\hline
\hline
 & $ \Omega $\,($ ^{\circ} $) & $ \omega $\,($ ^{\circ} $) & $ \nu $\,($ ^{\circ} $) \\
\hline
SC1 & 209.433\,009 & \hphantom{0}\hphantom{0}0.980\,870 & \hphantom{0}84.729\,131 \\
SC2 & 209.430\,454 & 205.692\,143 & 359.976\,125 \\
SC3 & 209.438\,226 & \hphantom{0}\hphantom{0}0.061\,831 & 325.619\,846 \\
\end{tabular}
\end{ruledtabular}
\end{table}

\begin{table}[ht]
\caption{\label{tab:eclipse_lun} The predicted Moon eclipses from May 2034 to May 2039 for the three TianQin satellites using the orbits of Table \ref{tab:initial}. The events all take place outside the 3+3 month observation windows (08 Jun to 06 Sep, 07 Dec to 07 Mar) with margins. }
\begin{ruledtabular}
\begin{tabular}{ccccc}
Start time (UTC)  & Type  & Duration & Satellite \\ 
\hline
08 Apr 2035 15:22:31 & Partial & 41 min & SC2 \\
27 Mar 2036 07:47:45 & Partial & 46 min & SC1 \\
25 May 2036 02:56:52 & Partial & 39 min & SC3 \\
24 Mar 2039 03:56:07 & Partial & 35 min & SC2 \\
\end{tabular}
\end{ruledtabular}
\end{table}

\section{\label{sec:conclusion} Concluding Remarks}

With the strategies of tuning the orbital radii and initial phases, Moon eclipses in the 3+3 month observation windows (08 Jun to 06 Sep, 07 Dec to 07 Mar) for TianQin can be avoided throughout the 5-year mission lifetime without designated avoidance maneuvers (orbit maintenance still required). Earth eclipses are unavoidable, but of much less concern as they only occur well outside the observation windows. As a viable candidate for TianQin, we propose using synodic resonant orbits at the ratio of approximately 1:8, i.e., the satellites completing 8 revolutions in one synodic month. The resonance significantly lowers the possibility of Moon eclipses. Accordingly, the radii near 100935 km have shown quite favorable performance in avoiding all eclipses during the observation windows for a 5-year mission starting in 2034. The permitted range of initial phases is $\sim 15^\circ$, granting a broad margin to orbit control. Note that the optimal radius may vary slightly around $\sim 100900$ km depending on the initial mission time selected, and that the eclipse-free periods can be extended at the expense of a reduced range of permitted initial phases. Our results will facilitate further trade studies in the orbit design and mission operations. The method presented here can be applied to other geocentric mission concepts as well. 

To help realize the proposed orbits, requirements on the delivery accuracy should be worked out. As suggested before, the most stringent requirement comes from the constellation stability, and not from the eclipse avoidance. The constellation stability performance is most sensitive to initial orbit errors in the satellites' radial positions and along-track velocities, and much less so to other errors, e.g. in cross-track directions, by 1-3 orders of magnitude. Preliminary estimates show that accuracies of $\sim 5$ m and $\sim 2$ mm/s (radial and along-track) may be needed to guarantee 3-month stability (e.g., the breathing angle variations within $\pm 0.1^\circ$). Thereby, various schemes, such as inter-satellite links, satellite laser ranging \cite{Mei2020}, GNSS (Global Navigation Satellite System), Chinese Deep Space Network, VLBI (Very-Long-Baseline Interferometry), etc., are now being considered to assess the orbit determination and control capabilities. More discussions are deferred to future work. 


\begin{acknowledgments}
The authors thank Yi-Ming Hu, Jian-dong Zhang, Haoying Zheng, Jianwei Mei, Defeng Gu, Jinxiu Zhang, Liang-Cheng Tu, and Jun Luo for helpful discussions and comments. Special thanks to Zhuangbin Tan for helping with the parallel computing, to the anonymous referees for valuable suggestions. The work is supported by the National Key R\&D Program of China (2020YFC2201202). XZ is supported by NSFC 11805287. 
\end{acknowledgments}






\bibliography{apssamp}

\end{document}